\documentclass[12pt]{iopart}
\usepackage{color}
\usepackage{epsfig}
\usepackage{rotating}

\newcommand{\nn}{\nonumber}
\newcommand{\beq}{\begin{equation}}
\newcommand{\eeq}{\end{equation}}
\newcommand{\bea}{\begin{eqnarray}}
\newcommand{\ben}{\begin{eqnarray*}}
\newcommand{\een}{\end{eqnarray*}}

\def\D0{D\O}

\begin{document}

\title{Charm production asymmetries from heavy-quark recombination}

\author{Thomas Mehen \footnote[1]{mehen@phy.duke.edu}}

\address{Department of Physics, Duke University, 
	Durham NC 27708, USA}

\date{\today}

\begin{abstract} 
Charm asymmetries in fixed-target hadroproduction experiments are sensitive to power corrections to the  QCD factorization
theorem for heavy quark production.  A power correction called  heavy-quark recombination has recently been proposed to
explain these asymmetries. In heavy-quark recombination,  a light quark or antiquark  participates in a hard scattering
which produces a charm-anticharm quark pair. The light quark or antiquark emerges from the scattering  with small momentum
in the  rest frame of the charm quark, and together they hadronize into a charm particle. The cross section for
this process  can be calculated within perturbative QCD up to an overall normalization. Heavy-quark recombination explains 
the observed $D$ meson and $\Lambda_c$ asymmetries with a minimal set of universal nonperturbative
parameters. 

\end{abstract}
In this talk I will discuss asymmetries in the production of charm particles in fixed-target 
hadroproduction experiments \cite{Aitala:1996hf,Aitala:2002uz,Adamovich:1998mu,Aitala:2000rd,Garcia:2001xj}.
These asymmetries are much larger than predicted by perturbative QCD
calculations, so they are a sensitive probe of  nonperturbative 
aspects of heavy particle production. Asymmetries in the production of light particles 
such as pions and kaons have also been observed, but  
the production of particles containing heavy quarks is under better theoretical control.
When the mass of the quark is heavier than $\Lambda_{\rm QCD}$, as is the case for charm, 
perturbative QCD can be applied even when the particle is produced without  
large transverse momentum. In addition, nonperturbative effects can be 
organized in an expansion in $\Lambda_{\rm QCD}/m_c$.  

The differential cross-section for the production of a charm particle, $H$, in the collision
of two hadrons, $A$ and $B$, is believed to factorize in the following manner \cite{Collins:1985gm}
\begin{eqnarray}\label{fact}
d \sigma [A+B &\rightarrow & H+X] = \\
&& \sum_{i,j} f_{i/A} \otimes f_{j/B} \otimes d\hat{\sigma}[ij \rightarrow c \bar c + X] \otimes
D_{c \rightarrow H} + ... \, . \nn
\end{eqnarray}
Here $i,j$ denote partons, $f_{i/A}$ and $f_{j/B}$ are parton distribution functions, $D_{c \rightarrow H}$ is the fragmentation
function for $c$ hadronizing into $H$ and $d\hat{\sigma}[ij \rightarrow c\bar c + X]$ is a perturbatively 
calculable short-distance cross section. The ellipsis represents  corrections to the factorized form of the cross section
that are suppressed by $\Lambda_{\rm QCD}/m_c$, or possibly $\Lambda_{\rm QCD}/p_\perp$ if $p_\perp \gg m_c$.

The leading contribution to the factorization theorem in Eq.~(\ref{fact}) predicts 
that charm particles and their antiparticles will be produced symmetrically. The leading 
order partonic processes, $gg\rightarrow c\bar c$ and $q \bar q \rightarrow c \bar c$,
produce charm quarks and anticharm quarks with identical kinematic distributions. The fragmentation functions 
$D_{c\rightarrow H}$ and $D_{\overline c \rightarrow \overline H}$ are identical due to the charge conjugation invariance of the 
strong interactions. No asymmetry between $H$ and $\overline H$ is generated at leading order in perturbation theory.
Next-to-leading order perturbative corrections 
\cite{Nason:1989zy,Beenakker:1988bq,Beenakker:1990ma,Frixione:1994nb} can generate asymmetries but these are quite small.
The asymmetry, defined by
\begin{eqnarray}
\alpha[H] = \frac{\sigma[H] - \sigma[\,\overline H \,]} {\sigma[H] + \sigma[\,\overline H \,]} \, ,
\end{eqnarray}
is never larger than a few percent.

The asymmetries observed in experiments can be much larger
\cite{Aitala:1996hf,Aitala:2002uz,Adamovich:1998mu,Aitala:2000rd,Garcia:2001xj}.  One well-known phenomenon  is the ``leading
particle  effect'' in fixed-target hadroproduction. Charm hadrons sharing a valence parton with the beam hadron  are produced in
greater numbers in the forward direction of the beam than particles that do not share a valence quark with the beam. For
instance, experiments with a $\pi^-$ beam incident on a nuclear target observe many more $D^-$ and $D^0$ than $D^+$ and
$\overline{D}^0$ in the forward direction of the $\pi^-$ beam. $\alpha[D^-]$ rises from nearly  zero in the central region, $x_F
\approx 0$, to 0.7 at the highest $x_F$ measured. In the  forward region  of the $\pi^-$ beam almost six $D^-$ are produced for
every $D^+$. Large asymmetries are also observed in charm baryon production. In $p N$ collisions $\alpha[\Lambda_c^+] \approx 1 $
for all $x_F > 0.2$.

Smaller asymmetries are also observed even when there is no leading particle effect. Examples include fixed-target
photoproduction~\cite{Anjos:1989bz,Alvarez:1993yb,Frabetti:1996vi}, where the beam has no valence quark quantum numbers, the
production of $D_s$ mesons in fixed-target hadroproduction experiments with $\pi$ or $p$ beams, and $\Lambda_c$ production in
experiments with $\pi^-$ beams. In the last case, there is no asymmetry from the leading particle effect because  the
$\Lambda_c^+$ shares a $d$ quark with the $\pi^-$ while the $\Lambda_c^-$ shares a $\bar u$ with the $\pi^-$. These
asymmetries are smaller than  those observed when there is a leading particle effect but still larger than
perturbative QCD predictions. 

Traditionally the asymmetries have been explained by  nonperturbative models of hadronization. A commonly used model is the
Lung string fragmentation model~\cite{Sjostrand:1986ys}  which can be implemented using PYTHIA~\cite{Sjostrand:2001yu}.
(Another model of string fragmentation can be found  in Ref.~\cite{Piskounova:2002gt}). The asymmetry is generated by the
``beam drag effect''  \cite{Norrbin:1998bw} in which the charm quark binds to the remnants of the incident hadron via the
formation of a color string. These models can be tuned to fit the data but this can require an unusually large charm quark mass
and large intrinsic transverse momentum for the partons in the incoming hadrons~\cite{Aitala:1996hf}. The PYTHIA Monte Carlo
with default parameters rarely  predicts the asymmetries correctly~\cite{Aitala:1996hf} and in the case of $\Lambda_c$
asymmetries in $\pi N$ collisions~\cite{Aitala:2000rd} gets the sign of the asymmetry wrong.  Another approach is the
recombination model  first introduced in Ref.~\cite{Hwa:1994uh} (for recent analyses using similar methods, see
Refs.~\cite{Rapp:2003wn,Likhoded:2000qc,Tashiro:kv}). In this model the charm quarks coalesce with spectator partons in the
beam hadrons whose momentum distribution is determined by double parton distributions. Finally there are
models that rely on the existence of intrinsic charm in the incident hadron wavefunction to generate the asymmetry 
\cite{Vogt:1995fs,Cuautle:1997ti,Herrera:1997qh}. These models are sensitive to a number of poorly determined nonperturbative
functions such as distributions of partons in the remnant and functions that parametrize recombination probabilities.

\begin{figure}[!t]
  \centerline{\epsfysize=6 truecm \epsfbox[100 550 350 700]{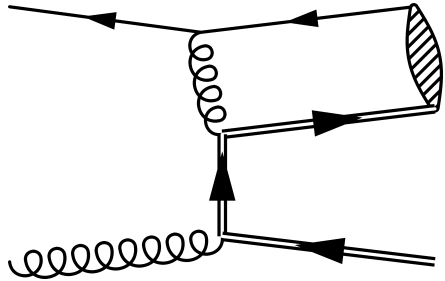}  }
 {\caption{Example of a diagram for $c\overline{q}$ recombination into a $D$ meson. Single lines are 
light quarks, double lines are heavy quarks and the shaded blob represents
hadronization into a state including the $D$ meson.}
\label{Fig1}}
\end{figure}

Since the asymmetries observed in nature are much larger than what is predicted by perturbative calculations, they are a
direct probe of the power corrections to the QCD factorization theorem. Recent work has demonstrated that an $O(\Lambda_{\rm QCD}/m_c)$  power correction called 
heavy-quark recombination provides a simple explanation of charm meson and baryon asymmetries in photo- and hadroproduction
experiments \cite{Braaten:2001bf,Braaten:2001uu,Braaten:2002yt,Braaten:2003vy}. This approach differs from previous 
nonperturbative models in that the asymmetry is generated in the short-distance process so cross sections
are calculable up to an overall normalization set by a few universal nonperturbative parameters.
For $D$ mesons, the dominant contribution comes from a process in which a light antiquark participates in a hard scattering process that produces a charm-anticharm
quark pair. The light antiquark emerges from the hard scattering with momentum of $O(\Lambda_{\rm QCD})$ in the rest frame of
the charm quark, then the light antiquark  and charm quark hadronize into a final state that includes a $D$ meson. The process
is depicted in Fig.~\ref{Fig1}.
This $c \overline q$ recombination process gives a contribution to the $D$ meson cross section
of the form 
\begin{eqnarray}
d{\hat\sigma}[\bar q g \rightarrow D +X] = \sum_n d{\hat\sigma}[\overline{q} g\rightarrow c\overline{q}(n) + \overline{c}]\,
\rho[c \overline{q}(n)  \rightarrow D] \, .
\end{eqnarray}
In this formula, $d{\hat\sigma}[\overline{q} g\rightarrow c\overline{q}(n) + \overline{c}]$ is a short-distance cross
section for producing a $c\overline{q}$ with quantum numbers denoted by $n$ and $\rho[c \overline{q}(n) \rightarrow D]$
parametrizes the hadronization of the $c \overline q$ into a state that includes a $D$ meson.

The NLO correction to the fragmentation cross section in Eq.~(\ref{fact}) includes
a subprocess $\bar q g\to \bar q c \bar c$ which is similar to the short-distance 
part of the  mechanism depicted in Fig.~\ref{Fig1}. However, in the NLO calculation
the $c$ and $\bar q$ hadronize independently, so this correction
does not account for the possiblility that the  $c\bar q$ can bind nonperturbatively and hadronize into the $D$.
For this reason, the heavy-quark recombination mechanism is an $O(\Lambda_{\rm QCD}/m_c)$ correction
to Eq.~(\ref{fact}) rather than part of the NLO correction to the fragmentation contribution.

Because the light
antiquark is massless it is natural to expect the heavy-quark recombination contribution to be a convolution of 
a short-distance cross section with a distribution
function that depends on the fraction of the light-cone momentum carried by the light quark. However, to lowest 
order in $\Lambda_{\rm QCD}/m_c$ only the leading moment of such a  distribution contributes. Because the
cross section is inclusive, the final state may include other soft quanta besides the $D$ meson.
Soft gluons emitted in the  hadronization process 
can change the total angular momentum 
and color quantum numbers of the $c\overline q$ produced in the short-distance process. Therefore the color and angular 
momentum quantum numbers of the
$c\overline q$ can be different from the $D$ meson. Amplitudes for production of $c \overline q$ in $L>0$ partial
waves are suppressed by  additional powers of $\Lambda_{\rm QCD}/m_c$ relative to S-waves and can be neglected.
It is also possible for the light antiquark flavor quantum number of the $D$ 
to be different than that of the $c \overline q$ due to light quark-antiquark pair production in the hadronization process. 
However,  light quark-antiquark pair production is suppressed in the large-$N_c$ limit of QCD so this effect is a
subleading correction to $D$ meson production. This leaves four parameters for heavy-quark recombination into $D^+$ mesons:
\begin{eqnarray}\label{para}
\rho_1       = \rho[c\bar{d}(^1S_0^{(1)})\rightarrow D^+],\quad 
\tilde \rho_1 = \rho[c\bar{d}(^3S_1^{(1)}) \rightarrow D^+], 
\\
\rho_8       = \rho[c\bar{d}(^1S_0^{(8)})\rightarrow D^+],\quad 
\tilde \rho_8 = \rho[c\bar{d}(^3S_1^{(8)}) \rightarrow D^+]. \nn
\end{eqnarray}
These parameters scale with the heavy quark mass as $\Lambda_{\rm QCD}/m_c$ and explicit 
expressions in terms of nonperturbative QCD matrix elements can be found  in
Ref.~\cite{Chang:2003ag}. Analogous parameters for $D^0$ and $D^-$ mesons are
obtained  by using isospin symmetry and  charge conjugation invariance,  while parameters for $D^{*+}$ states are related to
those in  Eq.~(\ref{para}) by heavy-quark spin symmetry.   
\begin{figure}[!t]
\centerline{\epsfysize=7.0truecm \epsfbox[100 550 350 700]{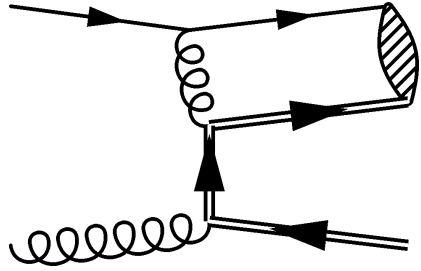} } 
\vspace{-0.5 in}{\caption{Example of a diagram for $c q$
recombination, the leading mechanism for charm baryon production via heavy-quark recombination.} 
\label{DiQ} }
\end{figure} 

To form a baryon from $c\overline q$ recombination requires the production of at least two light quark-antiquark
pairs which is suppressed by $1/N_c^2$, so $c \overline{q}$ recombination should contribute mostly to $D$ meson
production. The process which gives the leading contribution to charm baryon production is $c q$ recombination
which is shown in Fig.~\ref{DiQ}.
This is the same as $c \overline q$ recombination except a light quark participates in the recombination process instead
of a light antiquark. A light quark-antiquark pair must be produced for the $c q$ to
hadronize into a state that has a charm baryon or charm meson, so there is a $1/N_c$ suppression in either case.
Thus, $c q$ recombination is the dominant heavy-quark recombination contribution to baryon production 
but is subleading for meson production. In this talk I will compare heavy-quark recombination predictions
with data on $\Lambda_c$ asymmetries. The heavy-quark
recombination contribution to the $\Lambda_c^+$ production cross section is  
\begin{eqnarray}\label{eta} 
%d{\hat\sigma}[q g\rightarrow \Lambda_c^+ + X] &=&  \\
%\sum_n && \!\!\!\!\!\!\!\!\!\! d{\hat\sigma}[q g\rightarrow cq(n) + \bar{c}]\,
%\eta[cq(n) \rightarrow \Lambda_c^+] \, .\nonumber
d{\hat\sigma}[q g\rightarrow \Lambda_c^+ ] = \sum_n d{\hat\sigma}[q g\rightarrow cq(n) + \bar{c}\,]\,
\eta[cq(n) \rightarrow \Lambda_c^+] \, .
\end{eqnarray}
The $\eta$ parameters are similar to the $\rho$ parameters described earlier. There are two 
possible color states, $\bar 3$ and $6$, and two possible spin states contributing at this order, for 
a total of four parameters for $\Lambda_c^+$ production:
\begin{eqnarray}
\eta_3 = \eta[c u(^1S_0^{(\bar{3})})\rightarrow \Lambda_c^+], \quad 
\tilde{\eta}_3 = \eta[c u(^3S_1^{(\bar{3})})\rightarrow \Lambda_c^+], \\
\eta_6 = \eta[c u(^1S_0^{(6)})\rightarrow \Lambda_c^+],\quad 
\tilde{\eta}_6 = \eta[c u(^3S_1^{(6)})\rightarrow \Lambda_c^+] . \nn
\end{eqnarray}
Isospin symmetry requires $\eta[c u(n)\rightarrow \Lambda_c^+] = 
\eta[c d(n)\rightarrow \Lambda_c^+]$. These parameters also scale with the heavy quark mass as 
$\Lambda_{\rm QCD}/m_c$. 
 
Techniques for calculating the short-distance cross sections
$d{\hat\sigma}[\overline{q} g\rightarrow c\overline{q}(n) + \overline{c}]$ and 
$d{\hat\sigma}[q g\rightarrow c q(n) + \overline{c}]$ 
as well as explicit expressions can be found in Refs.~\cite{Braaten:2001bf,Braaten:2003vy}.
All these cross sections  are strongly peaked
in the forward direction of the initial light  quark or antiquark. Define
$\theta$ to be the angle between the light quark (or antiquark)
in the initial state and the  $c q$  (or $c\overline q$)
in the center-of-momentum frame. The ratio of the recombination cross section
to the dominant leading order fragmentation cross section, $gg\rightarrow c \overline c$,
at $\theta =\pi/2$ is 
\begin{eqnarray}
\left.{d \hat{\sigma}[\overline{q}g \rightarrow c\overline{q}(n) +\overline{c}]  \over 
d \hat{\sigma}[g g\rightarrow \overline{c} c]}\right|_{\theta = \pi/2}
\sim 
\left.{d \hat{\sigma}[q g \rightarrow c q(n) +\overline{c}]  \over 
d \hat{\sigma}[g g\rightarrow \overline{c} c]}\right|_{\theta = \pi/2} 
\sim   \alpha_s {m_c^2 \over \hat{s}} \, ,
\end{eqnarray}
where $\hat{s}$ is the parton center-of-mass energy squared. When the $cq$ or $c \overline q$
emerges at right angles to the incoming partons, heavy-quark recombination is suppressed relative to 
fragmentation by a kinematic factor of $m_c^2/\hat{s}$.
It is also suppressed when the charm quark is produced in the backward direction, $\theta =\pi$:
\begin{eqnarray}
\left.{d \hat{\sigma}[\overline{q}g \rightarrow c\overline{q}(^3S_1^{(i)}) +\overline{c}]  \over 
d \hat{\sigma}[g g\rightarrow \overline{c} c]}\right|_{\theta = \pi}
\sim 
\left.{d \hat{\sigma}[q g \rightarrow c q(^3S_1^{(i)}) +\overline{c}]  \over 
d \hat{\sigma}[g g\rightarrow \overline{c} c]}\right|_{\theta = \pi}
&\sim&   \alpha_s {m_c^2 \over \hat{s}}   \\
\left.{d \hat{\sigma}[\overline{q}g \rightarrow c\overline{q}(^1S_0^{(i)}) +\overline{c}]  \over 
d \hat{\sigma}[g g\rightarrow \overline{c} c]}\right|_{\theta = \pi}
\sim 
\left.{d \hat{\sigma}[q g \rightarrow c q(^1S_0^{(i)}) +\overline{c}]  \over 
d \hat{\sigma}[g g\rightarrow \overline{c} c]}\right|_{\theta = \pi}
&\sim&   \alpha_s {m_c^6 \over \hat{s}^3}   \,.
\end{eqnarray}
Note that the backward suppression is greater when the $c q$ or $c\overline q$ is produced in a $^1S_0$ state.
In the forward direction, there is no kinematic suppression:
\begin{eqnarray}
\left.{d \hat{\sigma}[\overline{q}g \rightarrow c\overline{q}(n) +\overline{c}]  \over 
d \hat{\sigma}[g g\rightarrow \overline{c} c]}\right|_{\theta = 0}
\sim 
\left.{d \hat{\sigma}[q g \rightarrow c q(n) +\overline{c}]  \over 
d \hat{\sigma}[g g\rightarrow \overline{c} c]}\right|_{\theta = 0} 
\sim   \alpha_s\, .
\end{eqnarray}
Heavy-quark recombination preferentially produces charm mesons and baryons 
in the forward direction of the incident light quark (or antiquark). The cross section is larger 
for charm particles that share a valence quark with one of the colliding hadrons because the structure 
functions of the valence quarks is largest. Thus, the heavy-quark recombination mechanism provides a natural 
explanation of the leading particle effect.

Another mechanism by which charm hadrons can be produced is by ordinary fragmentation 
of charm quarks produced in $\overline c q$ or $\overline c \, \overline q$ recombination. This process,
called ``opposite side recombination'', gives the following contributions to charm hadron 
 production:
\begin{eqnarray}
\label{osr-q}
d{\hat\sigma}[qg\rightarrow  H + X] &=& 
\sum_{n,\overline{D}} d{\hat \sigma}[q g\rightarrow \overline{c}q(n)+c]\, \rho[\overline{c}q(n) \rightarrow \overline{D}] 
\otimes D_{c \rightarrow H} \, ,\\ 
\label{osr-qbar}
d{\hat\sigma}[\overline q g\rightarrow H + X] &=& \sum_{n,\overline B} 
d{\hat\sigma}[\overline{q} g\rightarrow \overline{c}\, \overline{q} (n) + c]\,  \eta[ \overline{c} \, \overline{q} (n) \to 
 \overline B \,] \otimes D_{c\rightarrow H}\, .
\end{eqnarray}
Here $\overline B$ is a charm antibaryon. The opposite side recombination mechanism
can generate asymmetries even when there is no leading particle effect. 

\begin{figure}[!t]  
  \centerline{\hspace{0.49 in}\epsfysize=10truecm
  \begin{turn}{270} 
    \epsfbox[14 14 627 807]{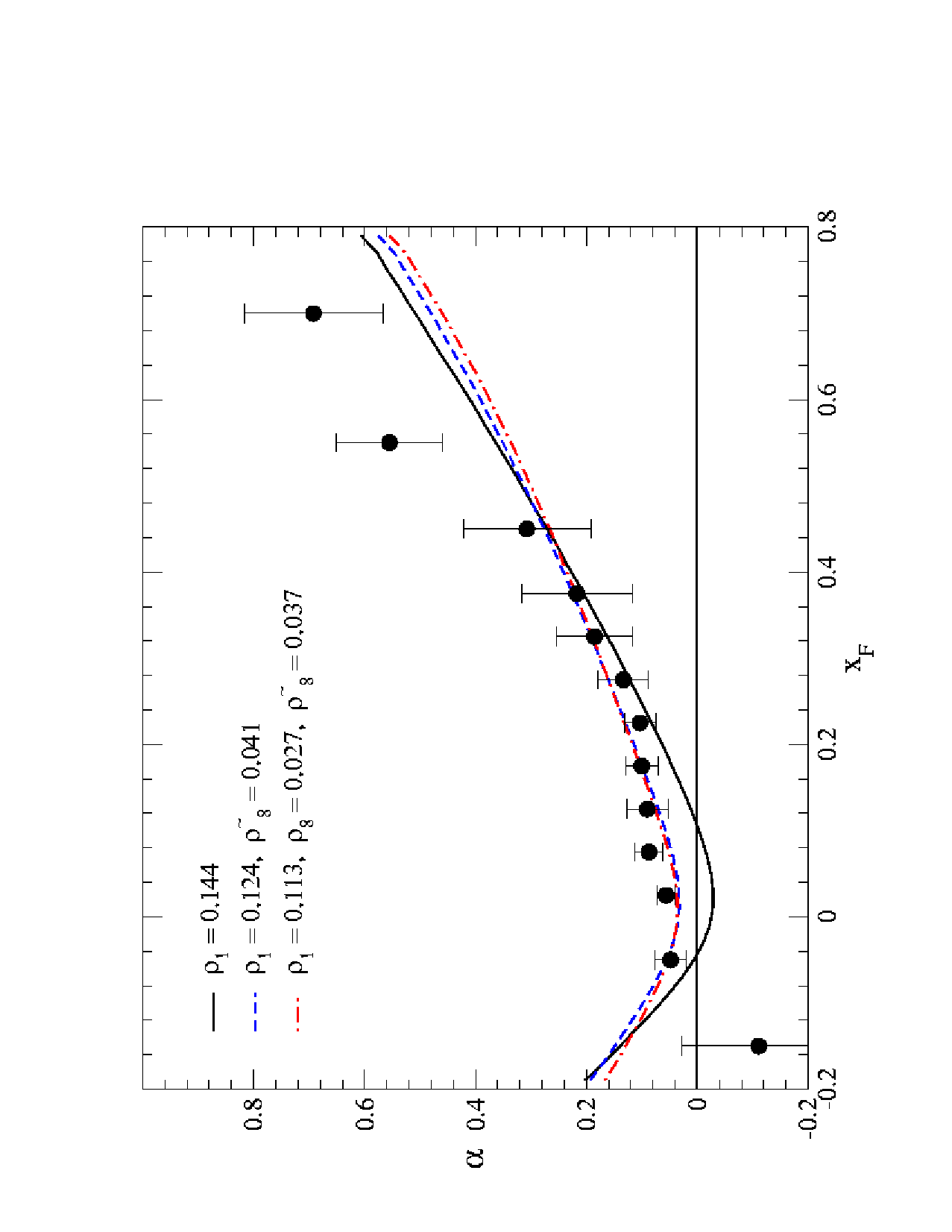}   
  \end{turn} }
  \centerline{\epsfysize=8truecm 
  \begin{turn}{270} 
    \epsfbox[14 14 545 650]{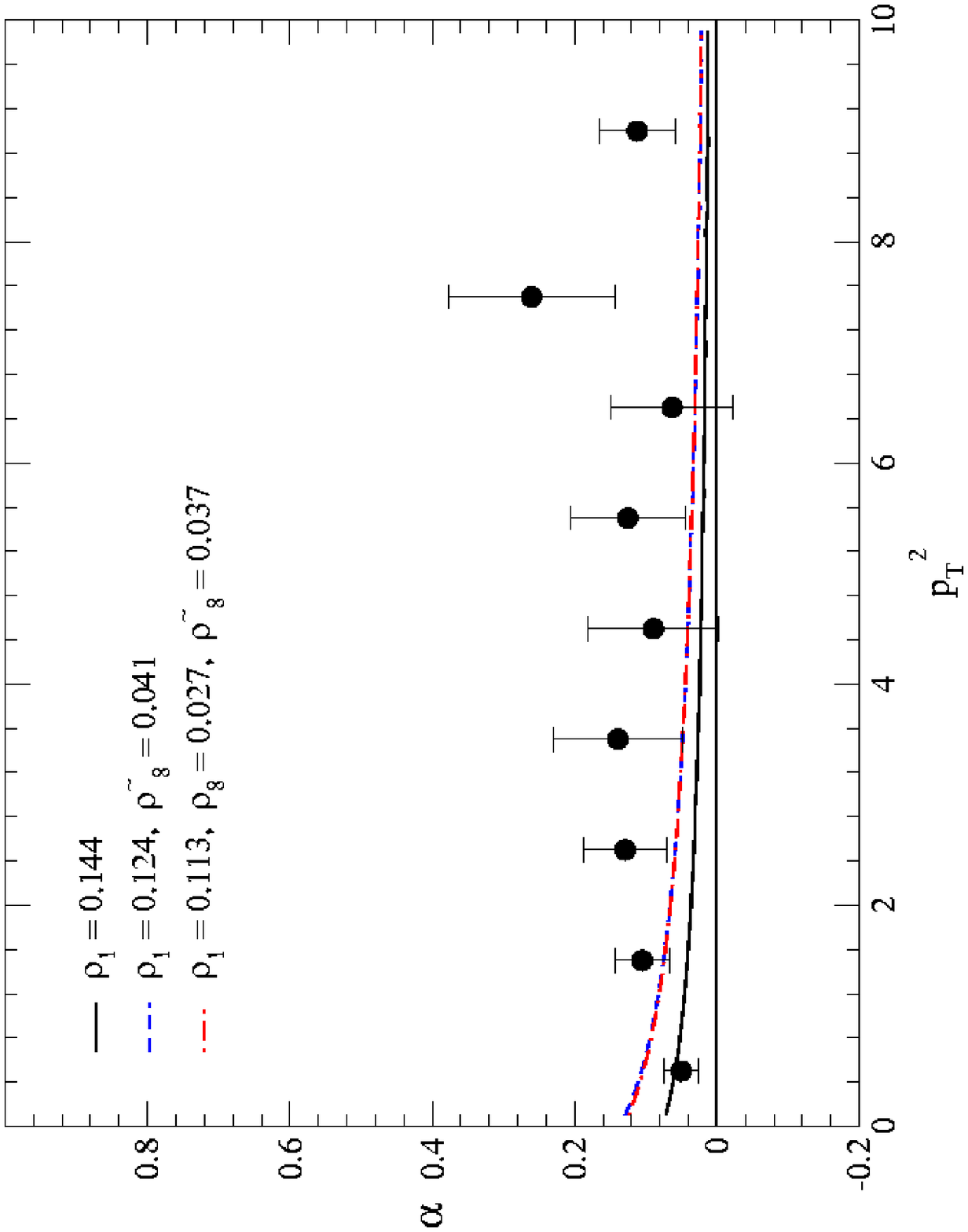}
  \end{turn} }
{ \caption{$\alpha[D^-]$ vs. $x_F$ and $p_\perp^2$ for a 500 GeV $\pi^-$ beam on a nuclear 
target \cite{Aitala:2002uz}. One-, two- and three-parameter fits  are solid, dashed and dot-dashed curves, respectively.} 
 \label{dminus}}
\end{figure}
Fig.~\ref{dminus} compares calculations of the $D^\pm$ asymmetry using the heavy-quark recombination mechanism 
with data from the E791 experiment in which a 500 GeV $\pi^-$ beam is incident on a nuclear target 
\cite{Aitala:2002uz}. In this analysis
the feeddown from $D^*$ mesons is included but feeddown from  excited $D$ meson states is neglected.
The charm quark mass is set equal to $1.5 \,{\rm GeV}$ and the renormalization and factorization
scales are $\sqrt{p_\perp^2+m_c^2}$. The parton distributions are GRV 98 LO \cite{Gluck:1998xa} for
the nucleon and GRV-P LO \cite{Gluck:1991ey} for the pion. The strong coupling constant
used is the one-loop expression for $\alpha_s$ with 4 active flavors and $\Lambda_{\rm
QCD}= 200\,{\rm MeV}$. Only the LO fragmentation diagrams are included in the calculation.
The effect of NLO perturbative corrections can be taken into account by multiplying the LO cross section by a K factor of
about $2$ \cite{Frixione:1994nb}.  The fragmentation functions  are $\delta$-functions times fragmentation 
probabilities taken from Ref.~\cite{Gladilin:1999pj}. At fixed-target energies
the single particle inclusive distributions $d\sigma/d x_F$ and $d\sigma/d p_\perp^2$ are better reproduced with
$\delta$-function fragmentation functions rather than, for example, Petersen fragmentation functions \cite{Frixione:1994nb}.

In this calculation of the $D$ meson asymmetry  the opposite side $\overline c \,\overline q$ recombination contribution of
Eq.~(\ref{osr-qbar}) is not included, so the result is independent of the $\eta$ parameters. The $\rho$ parameters were
determined by a global analysis of all measurements of $D$ meson asymmetries in fixed-target hadroproduction experiments
\cite{bkjm}. Fits with one, two and three parameters were performed. (A four-parameter fit did not yield significantly better
results than the three-parameter fit.)  The results of all three fits are shown in Fig.~\ref{dminus}. Note that even a fit
with one parameter, $\rho_1 = 0.14$ and all other $\rho$ parameters set to zero, does a good job of describing the data in
the forward region where heavy-quark recombination is most important. Inclusion of other parameters is necessary to obtain
agreement in the central region $x_F \approx 0$. 
\begin{figure}[!t]
  \centerline{\hspace{0.75 in}\epsfysize=10.0truecm
  \begin{turn}{270} 
    \epsfbox[14 14 627 807]{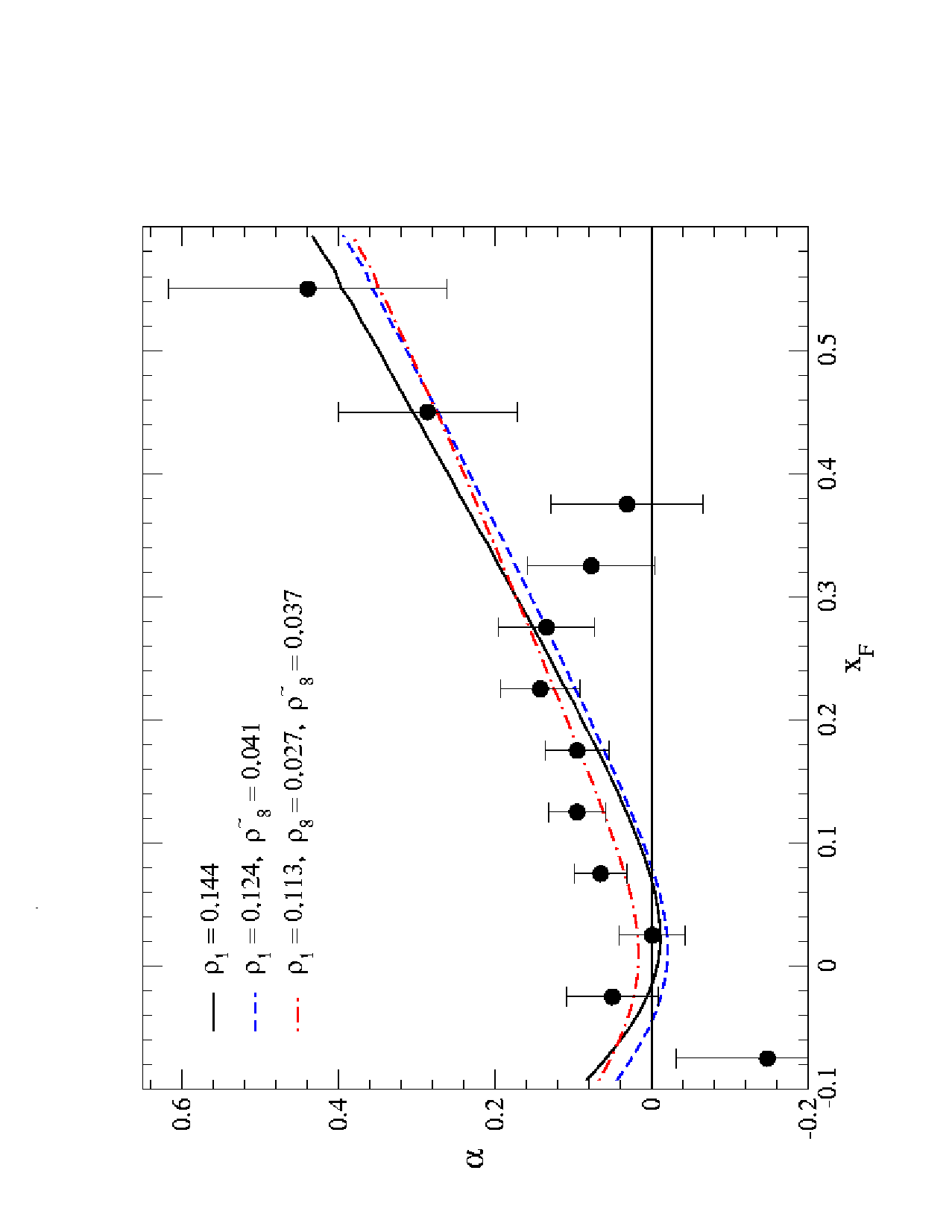} \end{turn}  }
{\caption{$\alpha[D^{*-}]$ vs. $x_F$ for a 500 GeV $\pi^-$ beam \cite{Aitala:2002uz}.
 Solid, dashed and dot-dashed curves are the same as Fig.~\ref{dminus}. }
 \label{dstar}} 
\end{figure} 
\begin{figure}[!t]
  \centerline{\hspace{0.75 in}\epsfysize=10.0truecm
  \begin{turn}{270} 
    \epsfbox[14 14 627 807]{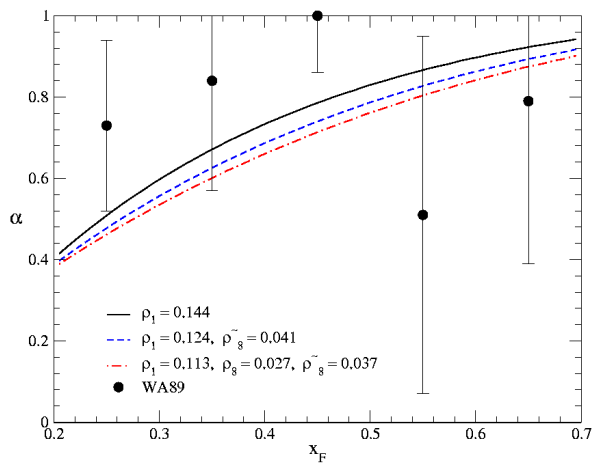}
  \end{turn}  }
{\caption{$\alpha[D_s^-]$ vs. $x_F$ for a 340 GeV $\Sigma^-$ beam \cite{Adamovich:1998mu}. 
Solid, dashed and dot-dashed curves are the same as Fig.~\ref{dminus}.} 
 \label{ds}} 
\end{figure}

The universality of the $\rho$ parameters is tested by looking at asymmetries of other $D$ mesons as well as asymmetries in
experiments with different beams. Fig.~\ref{dstar} compares the predictions of the heavy-quark recombination mechanism for
$\alpha[D^{*-}]$ with data from the E791 experiment \cite{Aitala:2002uz}. Fig.~\ref{ds} shows data and a calculation  of
$\alpha[D_s^-]$ for the WA89 experiment which has a 340 GeV $\Sigma^-$ beam incident on a nuclear target \cite{Adamovich:1998mu}.
These calculations use the same parameter sets as in Fig.~\ref{dminus} and the agreement with experiment is excellent.

\begin{figure}[!t]
\centerline{
\includegraphics*[width=7.77cm,angle=0,clip=true]{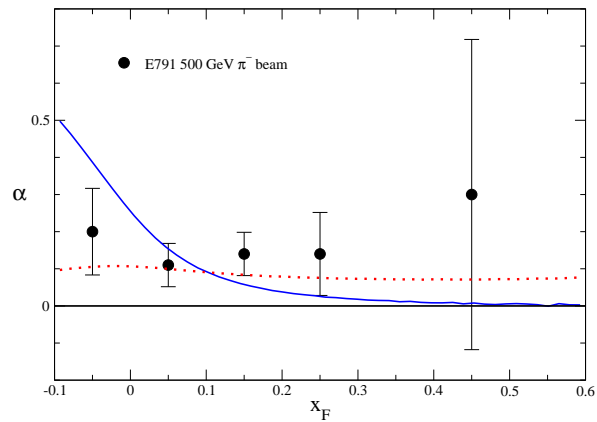}}
\caption{
 $\alpha[\Lambda_c^+]$ vs. $x_F$
for a 500 GeV $\pi^-$ beam \cite{Aitala:2000rd}.
The solid curve is the best single-parameter fit with $\eta_3 = 0.22$, 
while the dotted curve is in the absence of $cq$ recombination.}
\label{asym-pi}
\end{figure}

\begin{figure}[!t]
\centerline{
\includegraphics*[width=7.77cm,angle=0,clip=true]{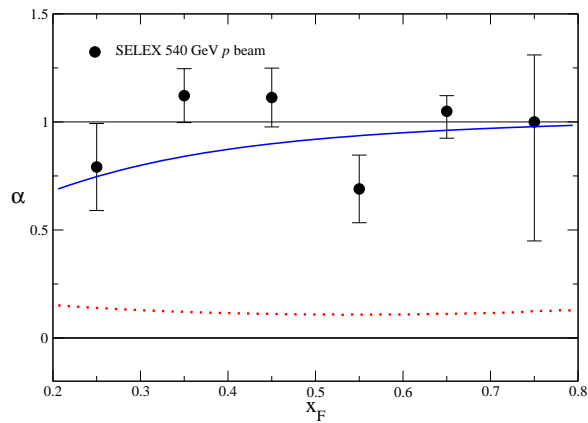}}
\caption{
$\alpha[\Lambda_c^+]$ vs. $x_F$
for a 540 GeV $p$ beam \cite{Garcia:2001xj}.
Fit parameters for the solid and dotted curves are the same as Fig.~\ref{asym-pi}.
The horizontal line at $\alpha = 1$ is the physical upper bound.}
\label{asym-p}
\end{figure}

Finally, the $c q$ recombination mechanism for charmed baryons
is tested by comparing to measurements of $\alpha[\Lambda_c^+]$ in experiments with 500 GeV
$\pi^-$ beams from E791 \cite{Aitala:2000rd} and 540 GeV $p$ beams from SELEX \cite{Garcia:2001xj}. The results are shown in Figs.~\ref{asym-pi} and \ref{asym-p}.
In the analysis of baryon asymmetries, the single nonvanishing $\rho$ parameter is chosen to be consistent with the
one-parameter fit to $D$ meson asymmetries. Setting all the $\eta$ parameters to zero gives the asymmetry
shown by the dotted lines 
in Figs.~\ref{asym-pi} and \ref{asym-p}, which is generated entirely by opposite side recombination. Though
this is adequate for the $\pi^-$ beam data, the $p$ beam data clearly requires an additional mechanism
for generating baryon asymmetries. A one parameter fit with $\eta_3 = 0.22$ is shown 
by the solid lines in  Figs.~\ref{asym-pi} and \ref{asym-p}. The
results of the calculation are in good agreement  with the baryon asymmetry measured in
both experiments. 

In this talk, I have shown that the heavy-quark recombination mechanism 
provides a natural mechanism for generating the observed 
charm meson and baryon asymmetries. There are four $c q$ recombination 
parameters and four $c \overline q$ recombination parameters which are
universal. The heavy-quark recombination mechanism   correctly accounts for 
observed $D$ meson and $\Lambda_c$ asymmetries. It is especially encouraging
that the same set of $\rho$ parameters can be used to correctly describe
$D$, $D^*$ and $D_s$ asymmetries in two different experiments with different beams.
$\Lambda_c$ asymmetries in both $\pi^-$ and $p$ beams are well described 
by a fit with a single $\rho$ parameter that is consistent with $D$ meson data
and a single $\eta$ parameter. In the future, a more detailed analysis 
of charm hadron asymmetries which thoroughly tests the universality of the $\rho$ and $\eta$
parameters should be performed. Once these universal parameters are determined,
 predictions for heavy-quark 
recombination cross sections for charm and bottom hadron production in a large 
variety of nuclear and particle physics experiments will be possible.

\section*{Acknowledgements}

It is a pleasure to acknowledge Eric Braaten, Yu Jia and Masaoki Kusunoki 
for collaboration on the work presented in this talk. I am especially
indebted to M. Kusunoki for help in preparing the figures. This research 
is partially supported in part by DOE Grants DE-FG02-96ER40945 and DE-AC05-84ER40150

\section*{References}
%\begin{references}

\end{document}